# Mach-Zehnder Interferometer for spin waves: computing applications.

Kirill A. Rivkin

Interferometers are some of the most important optical devices, yet their spin wave based analogues so far received limited attention. In this work we demonstrate how one can design Mach-Zehnder Interferometer (MZI) operating on spin waves, and use it to construct a spin wave version of a well known[1] MZI based AI computing device. Modeling, performed both micromagnetically and by using a novel, finite difference based method, shows that some of the achievable operational parameters are orders of magnitude superior to both optical and CMOS based implementations.

It has been demonstrated[2] that magnetic refractive index can be defined in a manner which leads to direct replication of optical equations such as Eikonal and Helmholtz (Eq.1). Value of refractive index can be controlled, for example, via inhomogeneous magnetic bias field and it is possible to replicate many optical mechanisms as "spin optics devices" with the identical functionality but operating on spin waves. In this article we analyze how this principle applies to one of the most important devices in optics – Mach-Zehnder interferometer.

To simplify the discussion we address in detail the case when the magnetic media is predominantly saturated via a strong uniform magnetic field $H_0$ applied in the out of plane direction, i.e. along the z axis. Micromagnetic modeling is performed via RKMAG code[3]. There are two magnetic layers, placed on top of each other (Figs. 3-4) with 10nm interlayer separation. The "bottom" layer is for spin wave propagation and its material parameters are representative of Yttrium Garnet[4] (YIG), $\mu_0 M_s = 0.18$ T, exchange stiffness $A = 3.65 \cdot 10^{-12}$ J/m, damping parameter $\beta = 0.0004$, thickness 15nm with discretization cell 15x15x15nm, i.e. below the exchange length. In order to suppress the boundary reflections in modeling there is 300nm wide layer along the edges where the damping gradually increases to 1. The top layer consists of hard granular FeCo film with $\mu_0 M_s = 1.25$ T, with average grain center to center distance of 7nm with 10% standard variation. Most of this layer is saturated in the direction parallel to the external bias field, except for a specific pattern (Figs. 3-4) where the magnetization is aligned in the opposite direction so it can generate in the propagation layer underneath it an additional bias field $\Delta H$. For the most modeling we assume no additional random material properties distribution, but when analyzing the computing precision in the end we will account for saturation magnetization (5% sigma) and anisotropy (10% sigma) distributions which for a standard recording head with 45nm track width result in 1.02nm recording transition sigma[5].

As established previously[2] Helmholtz-like equation for the magnetostatic potential $\varphi(x, y)$ is:

$$\left[\frac{\partial^2 \varphi(x,y)}{\partial x^2} + \frac{\partial^2 \varphi(x,y)}{\partial y^2}\right] + \acute{n}^2 k_0^2 \varphi(x, y) = 0 \qquad (1),$$

where some specific value of the bias field $H_0$ is selected as the "base" with a corresponding wavevector $k_0$ and a relative magnetic refractive index $\acute{n} = 1$, and for an arbitrary bias field $H_0 + \Delta H$ a relative magnetic refractive index is defined as $\acute{n}(x, y) = \frac{n(x,y)}{n_0}$. It can be obtained analytically as:

$$\acute{n}(x,y) = \frac{\sqrt{(1+\chi(H_0))}}{\sqrt{(1+\chi(H_0+\Delta H))}}, \qquad (2)$$

$$\chi = \frac{\omega_0 \omega_M}{\omega_0^2 - \omega^2}, \qquad \omega_M = \gamma\mu_0 M_s, \qquad \omega_0 = \gamma\mu_0(H_0 + \Delta H(x,y) - M_s),$$

where $M_s$ is saturation magnetization, $\gamma$ is gyromagnetic ratio, $\omega$ is the frequency at which spin waves are excited. To account for the exchange interaction[6] additional component $Dk^2$ proportional to wavevector amplitude squared can be added to $\omega_0$. To account for non-uniform magnetization one can add an additional contribution to $\acute{n}$ in (Eq.2) proportional to $i\nabla \cdot \boldsymbol{M}$. Similar expressions exist for the in plane saturated case, with an important difference that the magnetic media becomes birefringent.

(Eq. 1) can support plane wave-like solutions when $\acute{n}$ is real, i.e. $(1 + \chi) < 0$, and evanescent waves when $\acute{n}$ is imaginary which can occur when $\Delta H$ is so large that even the uniform mode has a higher frequency compared to $\omega$. In a more general case $\acute{n}$ can be a function of sample's geometry, saturation magnetization, temperature. While spin optics devices can be constructed by altering the corresponding parameters, in the present article we consider using the bias field only due to the relative simplicity with which it can be manipulated.

In order to connect micromagnetic modeling with (Eq.1) we need to create a database of $\acute{n}(\Delta H)$. It can be done analytically[2] (Eq.2) or by micromagnetically modeling a simple waveguide structure. It is created by subjecting a sample to a combination of a uniform field $H_0$ =187 kA/m and an additional "cladding" field 87KA/m which on both sides surrounds the 300nm wide waveguide "core": for the given frequency (Fig. 1) spin waves propagate in the center but are reflected from the "cladding" where the field is too high. The value of $H_0$ chosen in this modeling experiment matches the effective magnetic field in the areas where spin waves propagate in more complex devices we are about to consider (Figs.3-4). Calculation of real and imaginary portions of $\acute{n}$ is accomplished (Fig.2) by varying the bias field within a rectangular segment within the waveguide's core (red rectangle in Fig.1) and observing the resulting changes in spin wave phase and amplitude.

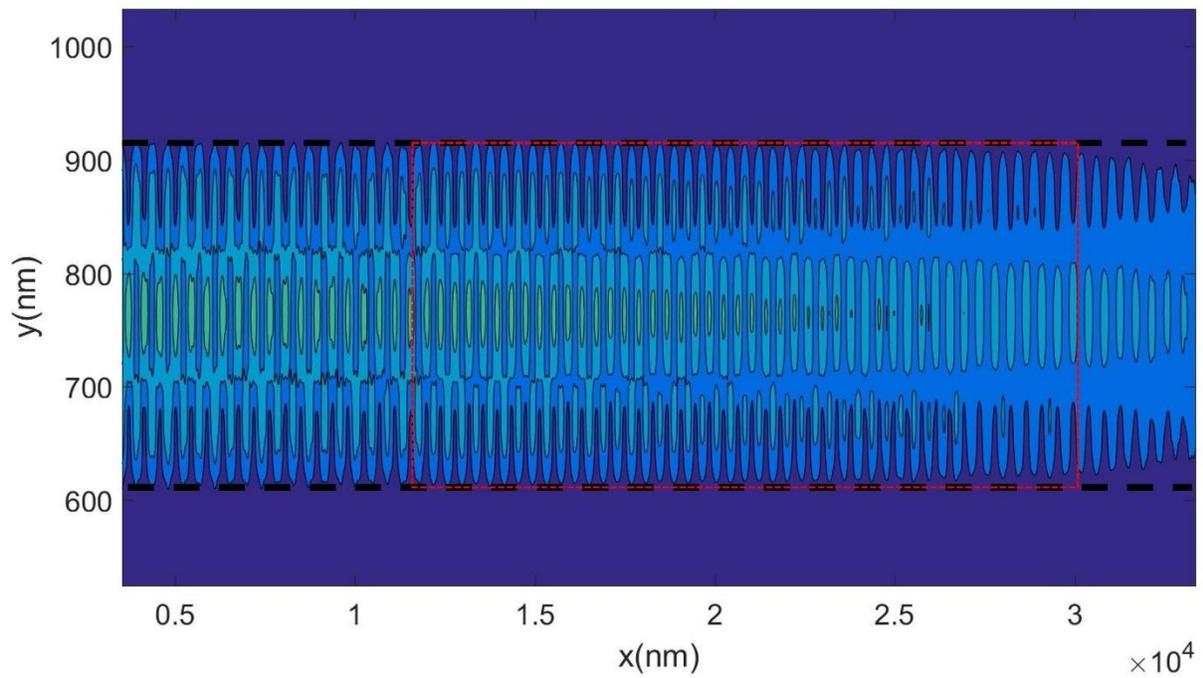

**Figure 1.** Magnetic waveguide with the core (the area inside black dashed lines) subjected to the out-of-plane bias field $H_0$ =187 kA/m, surrounded by the "cladding" where the total bias field is increased to 274 KA/m. To determine the effective refractive index the oscillation amplitude (shown here) and phase are modeled for an additional field (32 kA/m in this particular case) being applied inside the red rectangle. As evident from a significant amplitude decay the refractive index here has a large imaginary part (Fig. 2).

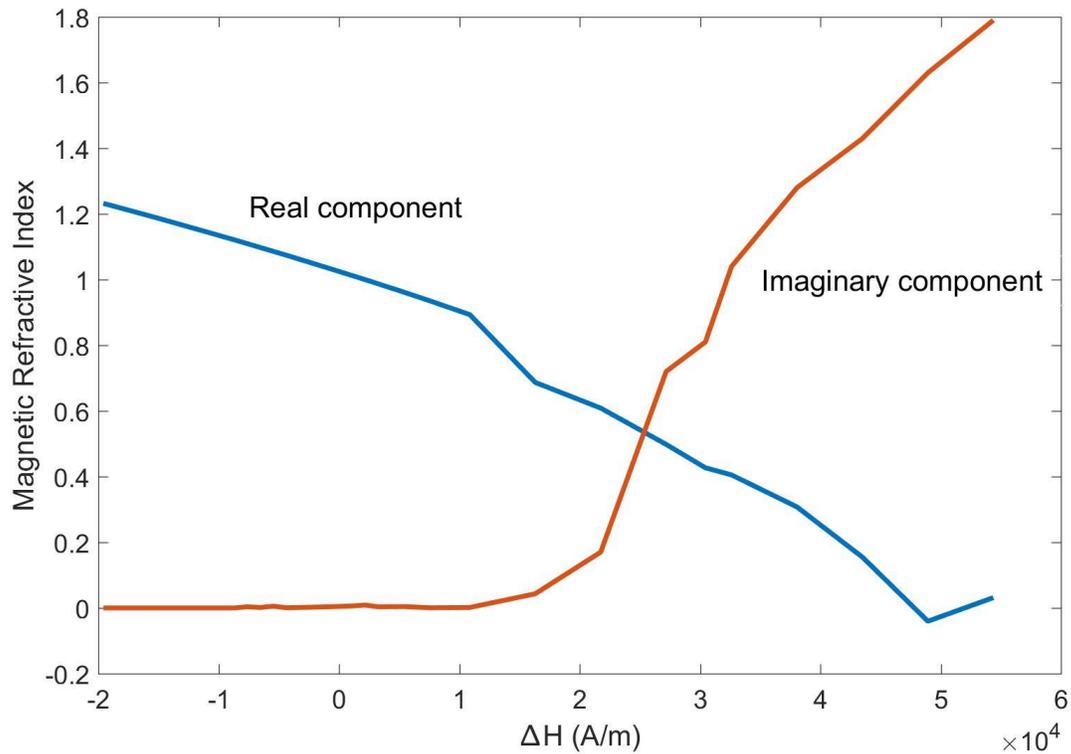

**Figure 2.** Magnetic refractive index estimated numerically for the setup shown in (Fig. 1) as a function of the additional bias field applied.

One can argue that if the field is negative and large enough to affect the equilibrium magnetization, than (Eq.1) is inapplicable, but in zeroth order the impact can still be simulated by using a magnetic refractive index with a large imaginary component. However, even when the equilibrium magnetization is not affected there is a difference between numerically (Fig.2) and analytically (Eq.2) obtained values, attributable[2] to differences between the underlining micromagnetic and magnetostatic assumptions. If one's goal is to accurately match micromagnetic results then using numerically obtained values (Fig. 2) is preferred.

Consider a directional coupler[7] – an important optical device where two waveguides are brought in proximity of one another allowing for the modes to travel between them. Its spin wave analogue (Fig. 3) can be constructed by recording a pattern in the hard layer where the negative polarity imitates the desired waveguide layout. "Cladding" in this case is due to the portions of the pattern with positive polarity (i.e. along $H_0$) generating enough bias field (Fig. 2) to provide areas with imaginary refractive index. Around the coupling segment the waveguides are separated by only 15nm wide strip with a positive polarity, whose thickness is insufficient to prevent the spin waves from interacting and whose length (i.e. the "coupler's length") determines the coupling amplitude (Fig.3.b), defined as the ratio of the output amplitudes in the upper and lower waveguide respectively.

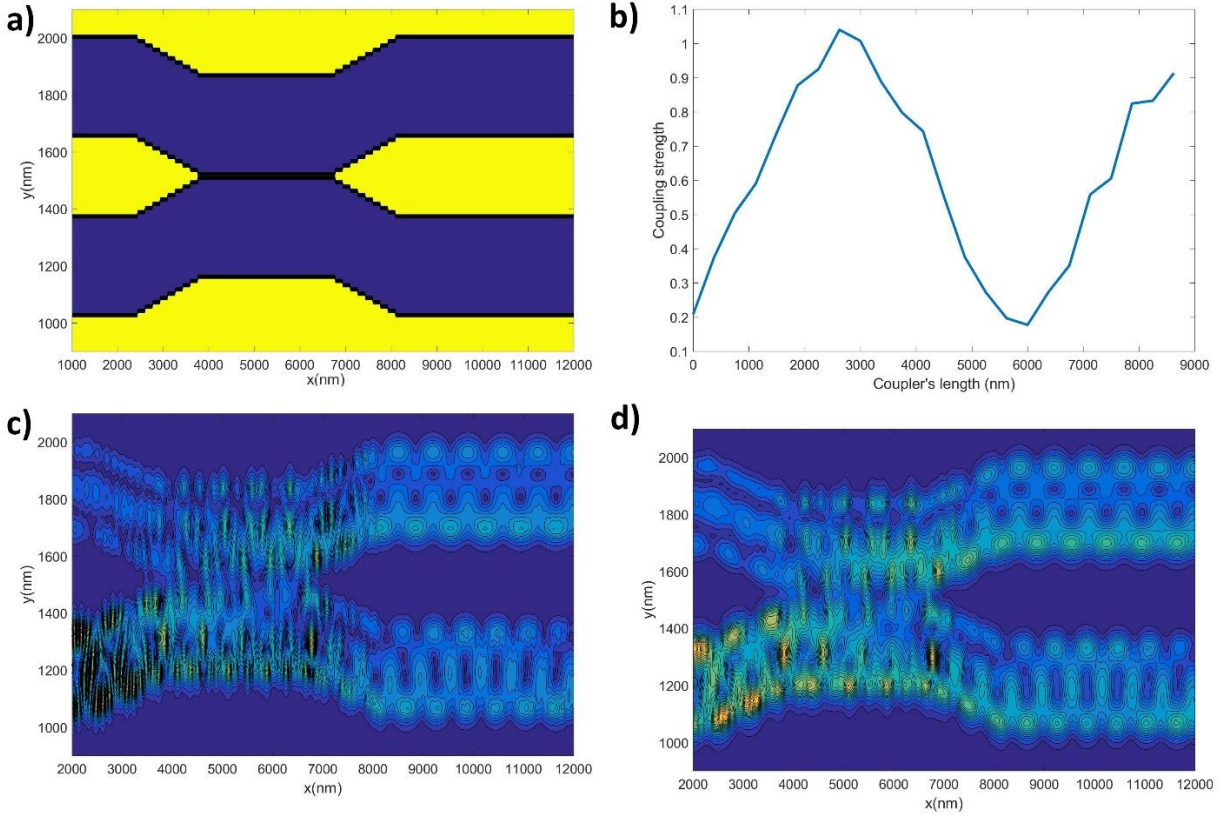

**Figure 3.** Directional coupler constructed using the magnetization pattern recorded onto hard layer (**a**), where blue color denotes the polarity opposite with respect to the external setting field $H_0 = 229.1\ kA/m$; spin wave is launched into the bottom waveguide and then partially crosses the gap between the waveguides. (**b**) Coupling amplitude as a function of the coupler's length, i.e. the section where waveguides are brought into direct proximity, optics-like sinusoid behavior is observed. (**c**) and (**d**) demonstrate spin wave oscillation amplitudes for the 3000nm coupler's length modeled using micromagnetics and finite difference method respectively.

Besides micromagnetics the system (Fig. 3) can be modeled by solving (Eq.1) using finite elements method, since out database of refractive index values (Fig. 2) for the given frequency allows one to compute $ń$ at every point. Alternatively (Eq.1) can be converted into optics-like phenomenological "time"-dependent equation:

$$\left[\frac{\partial^2 \varphi(x,y,t)}{\partial x^2} + \frac{\partial^2 \varphi(x,y,t)}{\partial y^2}\right] = \frac{ń^2}{c_0^2}\frac{\partial^2 \varphi(x,y,t)}{\partial t^2} \qquad (3)$$

where $c_0$ is spin wave speed when $\Delta H = 0$. Obviously spin waves do not follow the relationship $\omega = ck$ used to derive (Eq.3), resulting in large errors when calculating the propagation speed when $ń \gg 1$ or $ń \ll 1$. This is not a significant concern for waveguides and other systems where the majority of propagation is confined to the area where $ń = 1$. Time envelope approximation can be applied $\varphi(x,y,t) = \varphi(x,y)f(t)e^{-i\omega t}$ and assuming $f(t)$ varies slowly:

$$\left[\frac{\partial^2 \varphi(x,y)}{\partial x^2} + \frac{\partial^2 \varphi(x,y)}{\partial y^2}\right] + \acute{n}^2 k_0^2 \varphi(x,y) = -2ik_0 \frac{\acute{n}^2}{c_0 f(t)} \frac{df(t)}{dt} \varphi(x,y) \quad (4)$$

In steady state the right hand side vanishes and (Eq.1) is fulfilled, therefore solving (Eq.4) with a standard finite difference integration scheme (Fig.3.d) yields the results near identical to a much more cumbersome and time intensive micromagnetic modeling (Fig.3.a). It can be shown that (Eq.4) remains applicable even in the presence of geometric (gaps, boundaries) or other inhomogenuities: it is sufficient to calculate the resulting changes in the local bias field compared to a uniform and infinite media sample and then solve (Eq.4) for a corresponding spatial distribution of $\acute{n}$. It successfully reproduce the resonant spectra even for objects such as uniformly saturated nanodots.

For waveguides if backreflections are not a major concern the envelope approximation can also be applied to the spatial dependence, $\varphi(x,y,t) = \varphi_0(x,y,t)e^{i\mathbf{k}\cdot\mathbf{r}-i\omega t}$:

$$k_{x0}\left(\frac{\partial \varphi_0}{\partial x}\right) + k_{y0}\left(\frac{\partial \varphi_0}{\partial y}\right) = -\frac{\acute{n}^2 k_0}{c_0} \frac{\partial \varphi_0}{\partial t} \quad (5).$$

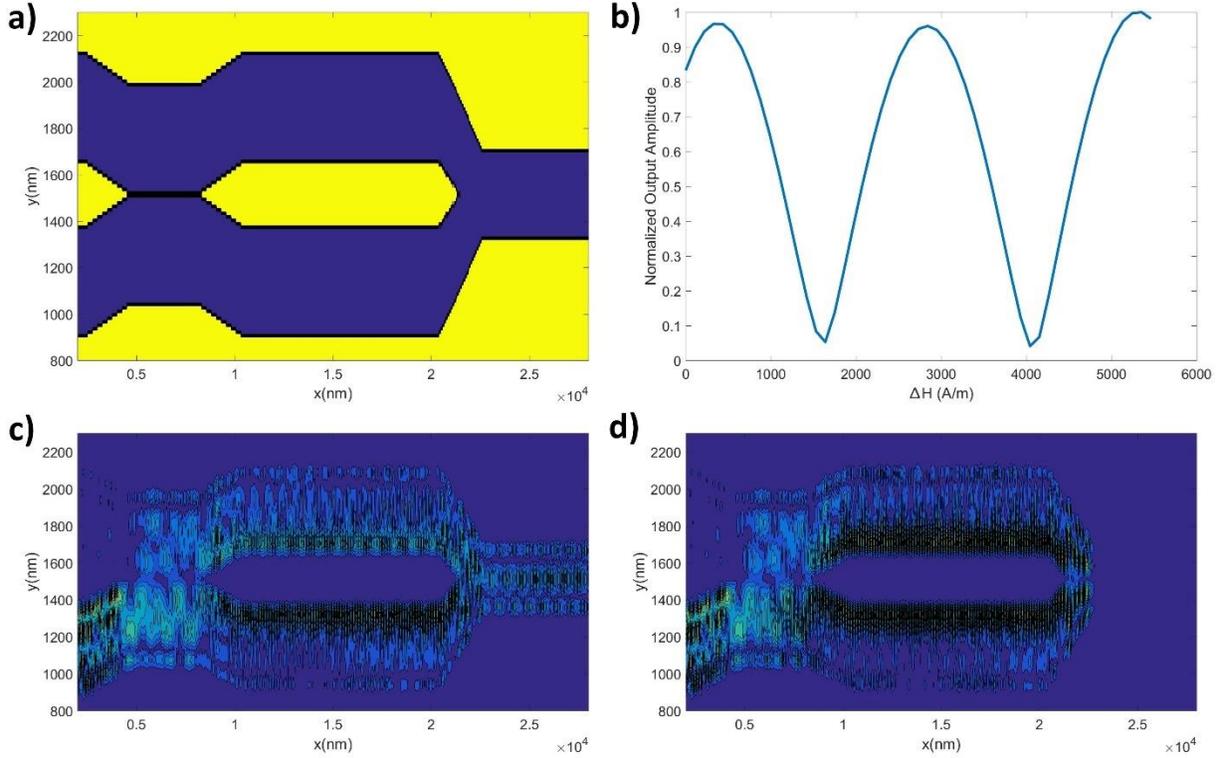

**Figure 4.** MZI where the input spin wave enters through the bottom waveguide, goes through the directional coupler, then subjected to the phase shift due to the additional bias field present in the bottom waveguide (the area inside the red dashed line) and then the two waveguides are joined into a single output. There is a uniform external field $H_0 = 229.1\ kA/m$ and the additional bias field is applied (**b**) in the lower waveguide in the opposite direction with respect to $H_0$. (**a**) depicts the corresponding magnetization pattern in the hard layer, (**b**) output amplitude as a function of the additional bias field, (**c**)

micromagnetically modeled oscillation amplitudes, for the in phase (zero additional bias field) and (**d**) out-of-phase (1.65 kA/m) cases.

We can create a proper Mach-Zehnder Interferometer (MZI) by first splitting the signal between the two waveguides (Fig. 4) via a coupler or by simply separating one waveguide into two, then subject one of the paths to some source of additional external field and then recombine the signals using another coupler or by joining the waveguides together. As in optics, there are plenty of practical applications. We can use it to measure small magnetic fields by shielding one of the waveguides while exposing the other to the external magnetic field, which is straightforward to do with predominantly inplane saturated geometry. With a typical sensitivity of magnetic refractive index to the external magnetic flux (Fig. 2) the two signals acquire a mutual $\pi$ phase shift within a distance of about 0.08-1.5 meters per nanotesla, depending on the value of the bias field $H_0$ and the operating frequency.

Another application is MZI based computing: the optics version is currently considered to be close to commercial realization[8]. Its basic functionality is a matrix-vector multiplication performed by first transforming it into a sequence of multiplications involving only two by two matrices and then encoding each one of them as operations whose combination was proven sufficient to represent an arbitrary matrix[9]: amplitude's attenuation, amplitude split between the waveguides and phase rotation. All three can be performed using an optical MZI element and an arbitrary matrix multiplier consists of many such elements. Rotation as a phase shift is produced by using a material with either temperature or electro-optical sensitive refractive index and is the only operation whose optical parameters can be adjusted at the execution time.

Consider how the same operations can be performed on the input signal vector $\begin{bmatrix} A \\ B \end{bmatrix}$ corresponding to the spin waves injected into the two input waveguides (Fig. 4):

$$\begin{bmatrix} \acute{A} \\ \acute{B} \end{bmatrix} = \begin{pmatrix} e^{i\phi_a} & 0 \\ 0 & e^{i\phi_b} \end{pmatrix} \begin{pmatrix} \alpha & 0 \\ 0 & \beta \end{pmatrix} \begin{pmatrix} a_{11} & a_{12} \\ a_{13} & a_{14} \end{pmatrix} \begin{bmatrix} A \\ B \end{bmatrix} \qquad (6)$$

Phase shift matrix applies additional phases $\phi_{a,b}$ by subjecting the respective waveguide (Fig.4) to the additional bias field generated by either an external current bearing wire (active method), or by a controlled reduction of saturation of the magnetization pattern recorded in the hard layer within the waveguide core (passive method).

Spin wave attenuator, i.e. multiplying the amplitude by a parameter $\alpha, \beta < 1$ is accomplished by using a coupler with a coupling strength $\frac{1-\alpha}{2}$ to redistribute the signal into an additional waveguide, apply there $\pi$ phase shift and consequently rejoin the signals (Fig. 4), which reduces the input signal by $\alpha$. Controllable attenuation is possible all the way to about 0.1-10% of the input value, depending on the waveguide design. Amplitude split matrix $\begin{pmatrix} a_{11} & a_{12} \\ a_{13} & a_{14} \end{pmatrix}$ in both optics and "spin optics" is accomplished via directional couplers (Fig. 3).

Both out-of-plane (Fig. 4) and inplane saturated (i.e. with much lower external field requirements) implementations have shown similar computing performance.

|  | Google TPU | Optics MZI | Spin Optics MZI |
|---|---|---|---|
| Computing Density TMAC/s/mm$^2$ | 0.58 | 0.56 | 1-10 |
| Energy/MAC, fJ | 430 | 30 | 0.01 (passive) – 2 (active) |
| Precision, bits | 8 | 5-6 | 5-12 |
| Latency, ps | 1400 | 100ps< | 1000-5000 |

**Table 1.** Performance parameters for various processor designs.

While MZI shown in (Fig. 4) illustrates each of these three functionalities the dimensions used are for demonstration purposes and are far from optimal for computing. First, the separation between the waveguides needs to be increased to at least 1-1.5um to make sure the wire field for active phase control impacts only one of the waveguides. Second, magnetic imperfections in the hard layer due to a combination of the recording process and material parameters distribution need to be taken into account. This creates two optimization scenarios. Key difference between them is precision, to which the phase shift operation is a minor contributor, most losses occurring in the joiner or the coupler. This can be offset by using a wide waveguide and therefore a long (>500nm) coupler so that random variations due to the recording quality mostly cancel each other and 1-5% maximum error is achievable, compatible with 5-7bit resolution. However, using a recordable magnetization pattern makes it possible to introduce changes in the finished product, canceling the errors due to material or design imperfections at the cost of relatively small (less than 0.1% of the total) changes in the recorded pattern. This can be shown to boost the resolution up to the maximum of 10-12bit. Precision is maintained by using a long phase shift area, which therefore requires small refractive index changes and produces less back-reflections. The joiner needs to be narrowed down (315nm in Fig. 4) since wider waveguides support multiple non-interacting modes, affecting the interference between the joined signals. Since in the zeroth order the equilibrium magnetization is not affected by the wire field as it only impacts the refractive index, the system's speed and latency are gated by spin wave propagation, which is about 9.5um/ns in (Fig.4). This design is large (about 25x2um) and slow, with a latency of about 3ns.

Alternative design focuses on speed at the cost of precision: by going to 5GHz operating frequency, shorter wavelength, narrower waveguide, shortened coupler and phase shift segments the dimensions can be reduced to 4x2um, resulting in a 4-5bit precision before the pattern correction, with 500ps latency.

In terms of energy consumption, the passive regime (phase shift via static magnetization pattern) is exceptionally low energy, but for the active regime, a copper wire with 1um cross-section is expected to consume between 0.5 fJ for fast/compact and 2fJ for slow/large chip design per operation.

Compared to conventional digital devices[10] (Tab. 1) either design is superior. Optical computer's speed and energy consumption is often gated not by the propagation but by how fast the refractive index can be altered. Still, it is an order of magnitude faster than the spin wave computer, but consumes much larger energy per operation.

Unique advantage of the proposed spin wave chip is that its manufacturing involves very basic wire lithography combined with recording a specific magnetization pattern, which can be completed in a matter of hours with negligible costs compared to multi-layered lithography of CMOS or high precision 3D printing for optical devices. Reprogramming an existing chip is equally simple.

Every other optical computing scheme can also be implemented with spin waves by using (Eq.4) to mimic its optical counterpart and in most cases the comparison of performance reveals the same ranking: spin wave implementation is orders of magnitude cheaper, more compact, with about two orders of magnitude lesser energy expenditure per operation, but it is an order of magnitude slower compared to the equivalent optical implementation.